\documentclass[times,number]{elsarticle}
\usepackage{amsmath}
\usepackage{amssymb}
\usepackage{graphics}
\usepackage{graphicx}
\usepackage{epsfig}
\usepackage{dcolumn}
\usepackage{bm}
\usepackage{lineno}

\begin{document}
\begin{frontmatter}

\title{A new type of RPC with very low resistive material}

\author[label1]{S. Chakraborty}
\author[label1]{S. Chatterjee}
\author[label1]{S.~Roy \corref{cor1}}
\ead{shreyaroy@jcbose.ac.in}
\author[label2]{A.~Roy}
\author[label1]{S.~Biswas\corref{cor}}
\ead{saikat@jcbose.ac.in, saikat.ino@gmail.com, saikat.biswas@cern.ch}
\author[label1]{S.~Das}
\author[label1]{S.~K.~Ghosh}
\author[label1]{S.~K.~Prasad}
\author[label1]{S.~Raha}

\cortext[cor]{Corresponding author}

\address[label1]{Department of Physics and Centre for Astroparticle Physics and Space Science
(CAPSS), Bose Institute, EN-80, Sector V, Kolkata-700091, India}
\address[label2]{Department of Particle Physics and Astrophysics, Weizmann Institute of Science, 7610001 Rehovot, Israel}

\begin{abstract}
There are several working groups that are currently working on high rate RPC's using different materials such as Si-based Ceramics, Low-resistive Glass, low-resistive bakelite etc. A new type of single gap RPC has been fabricated using very low-resistive carbon-loaded PTFE material to compete with all these other groups and materials. In terms of bulk resistivity, this material is the lowest and should in principle be able to work at the highest rates, provided the material can withstand working bias and radiation. The efficiency and noise rate of the RPC are measured with cosmic rays. The detail method of fabrication and first experimental results are presented. 
\end{abstract}
\end{frontmatter}


\section{Introduction}\label{intro}

Single gap Resistive Plate Chamber (RPC) is one of the most widely used detector technologies for trigger and tracking in high energy physics experiments for its excellent efficiency and time resolution \cite{RSRC81, SB109, SB309}. Rate handling capacity of RPC can be increased using low resistive electrode plate and operating the RPC in the avalanche mode. 

Triple GEM (Gas Electron Multiplier) detectors will be used in the first two stations of the CBM (Compressed Baryonic Matter) \cite{CBM} muon chamber (MUCH) at the future Facility for Antiproton and Ion Research (FAIR) \cite{FAIR} in Darmstadt, Germany. Given the interaction rate of 10 MHz the expected particle flux on the first station of CBM-MUCH will be about 3.1~MHz/cm$^2$. Maximum particle flux on the 3$^{rd}$ and 4$^{th}$ stations of the CBM-MUCH have been estimated to be 15~kHz/cm$^2$ and 5.6~kHz/cm$^2$, respectively, for central Au-Au collisions at 8 AGeV. We are exploring the possibility of using RPCs for the 3$^{rd}$ and 4$^{th}$ stations of CBM-MUCH. To obtain higher rate capability a prototype RPC using low resistive material has been built and tested in avalanche mode. The detailed method of building of the prototype detector, set-up, measurement and the first test results are presented in this article.


\section{Detector descriptions and experimental set-up}\label{setup}
We have designed a prototype RPC with a carbon-loaded Polytetrafluoroethylene (PTFE) material commonly known as Teflon. This particular sample is 25\% carbon-filled having a bulk resistivity of 10$^5$ $\Omega$-cm. The bulk resistivity has been measured using the method as described in Ref~\cite{KKM}. It is to be mentioned here that that carbon-loaded PTFE or any other carbon-loaded resistive plate can be tuned according to the resistivity requirement by changing the carbon-content and these resistive plates can be used for high rate RPC R\&D. The relationship between carbon-content and the resistivity is non-linear and substantial R\&D needs to be done on this front alone to find suitable materials.

In this detector two 15~cm~$\times$~15~cm plates of thickness 1~mm are used to build the chamber. 2~mm gas gap has been maintained using four 1~cm~$\times$~15~cm edge spacers and four button spacers of 1~cm diameter. Two gas nozzles are used for gas inlet and outlet. All the spacers and nozzles are made of polycarbonate. The measured surface resistivity of the carbon-loaded PTFE has been found to be 20~k$\Omega$/$\Box$. Since the surface resistivity of carbon-loaded PTFE is very low, the material has not been coated with graphite for high voltage distribution. As the surface of the material has been found to be smooth by visual inspection, we have not used oil coating in this case.

\begin{figure}[htb!]
\begin{center}
\includegraphics[scale=0.35]{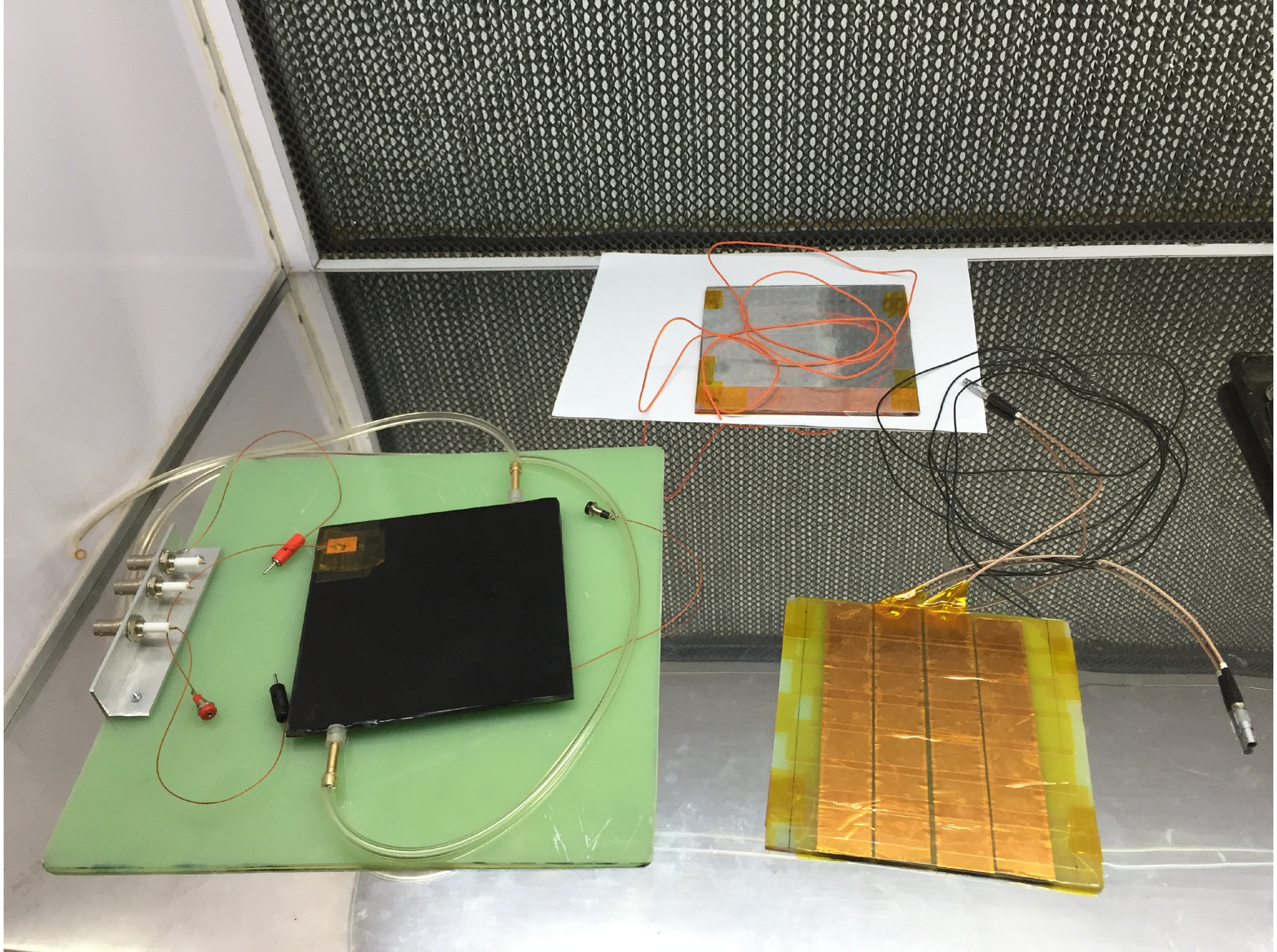}
\caption{Complete RPC along with copper pick-up strips.}
\label{complete}
 \end{center}
\end{figure}

100\% R-134a (Tetrafluoroethane) has been used as the sensitive gas for the chamber. The induced signal is readout by orthogonal pick-up strips placed on two sides of the chamber. There are four copper strips of dimension 2.5~cm~$\times$~15~cm with a separation of 2~mm, at the central part of the module in each side. The copper pick-up strips are pasted on a 3~mm thick G-10 board of dimension 15~cm~$\times$~15~cm. The ground plane of the pick-up panel has been made of aluminium. The complete RPC module along with the pick up strips are shown in Figure~\ref{complete}.


\begin{figure}[htb!]
\begin{center}
\includegraphics[scale=0.25]{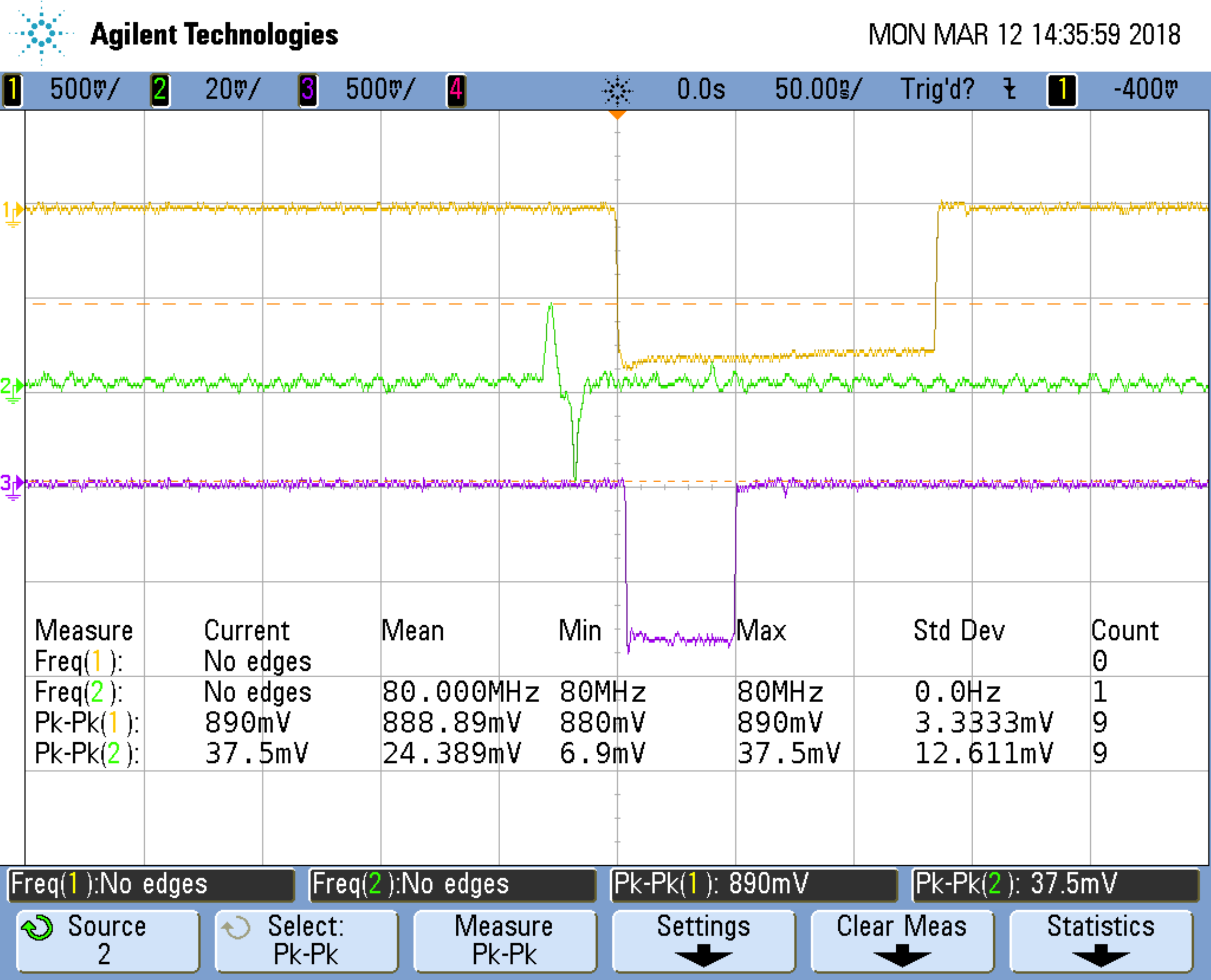}
\caption{Signal after the preamplifier.}
\label{signal}
\end{center}
\end{figure}


Differential voltage has been applied to the chamber to produce the electric field inside the gas gap. A charge sensitive preamplifier (VV50-2) \cite{Preamp} with gain 2~mV/fC and shaping time 300~ns has been used for the signal collection. In the Figure~\ref{signal} the Green line represents the signal after the preamplifier, which
has been taken from the positive plane of the detector. There was always a reflected negative part in each signal  because of impedance mismatch and this negative part has been used to discriminate the signal from noise using a leading edge discriminator (LED). For efficiency measurement  a cosmic ray trigger set-up is used keeping two plastic scintillator detectors of dimension 20~cm~$\times$~20~cm and 2~cm~$\times$~10~cm above the chamber and one with dimension 10~cm~$\times$~10~cm  below it. Effectively an overlap area of 2~cm~$\times$~10~cm is available for triggering purpose. The coincidence signals from these three scintillator detectors are taken as master trigger and shown in Figure~\ref{signal} (Yellow). The width of the trigger has been set at 120~ns. The discriminated RPC signal shown in magenta in Figure~\ref{signal} has been taken in coincidence with the master trigger to get a four-fold signal. The three-fold master trigger and the four-fold signals are counted using a NIM scaler. The un-triggered discriminated RPC signals are also counted to measure the noise rate.



\section{Results}\label{res}

\begin{figure}[htb!]
\begin{center}
\includegraphics[scale=0.5]{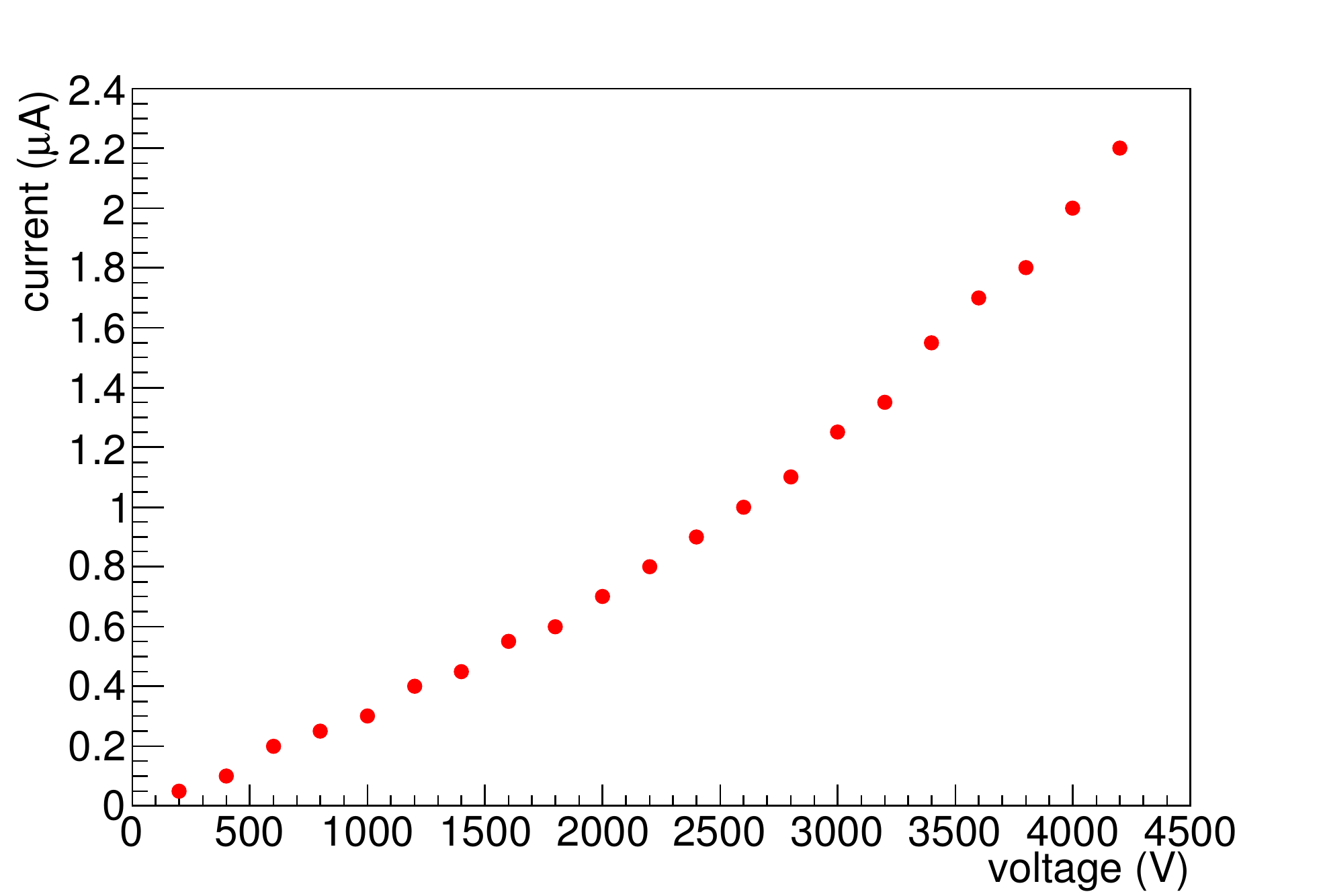}
\caption{V-I characteristics.}
\label{vi}
 \end{center}
\end{figure}

In this study the prototype chamber has been tested for V-I characteristics, variation of noise rate and efficiency as a function of applied voltage. Equal voltages of opposite polarity have been applied on two planes of the module and the leakage current has been measured. The V-I characteristics for the module is shown in Figure~\ref{vi}.


\begin{figure}[htb!]
\begin{center}
\includegraphics[scale=0.5]{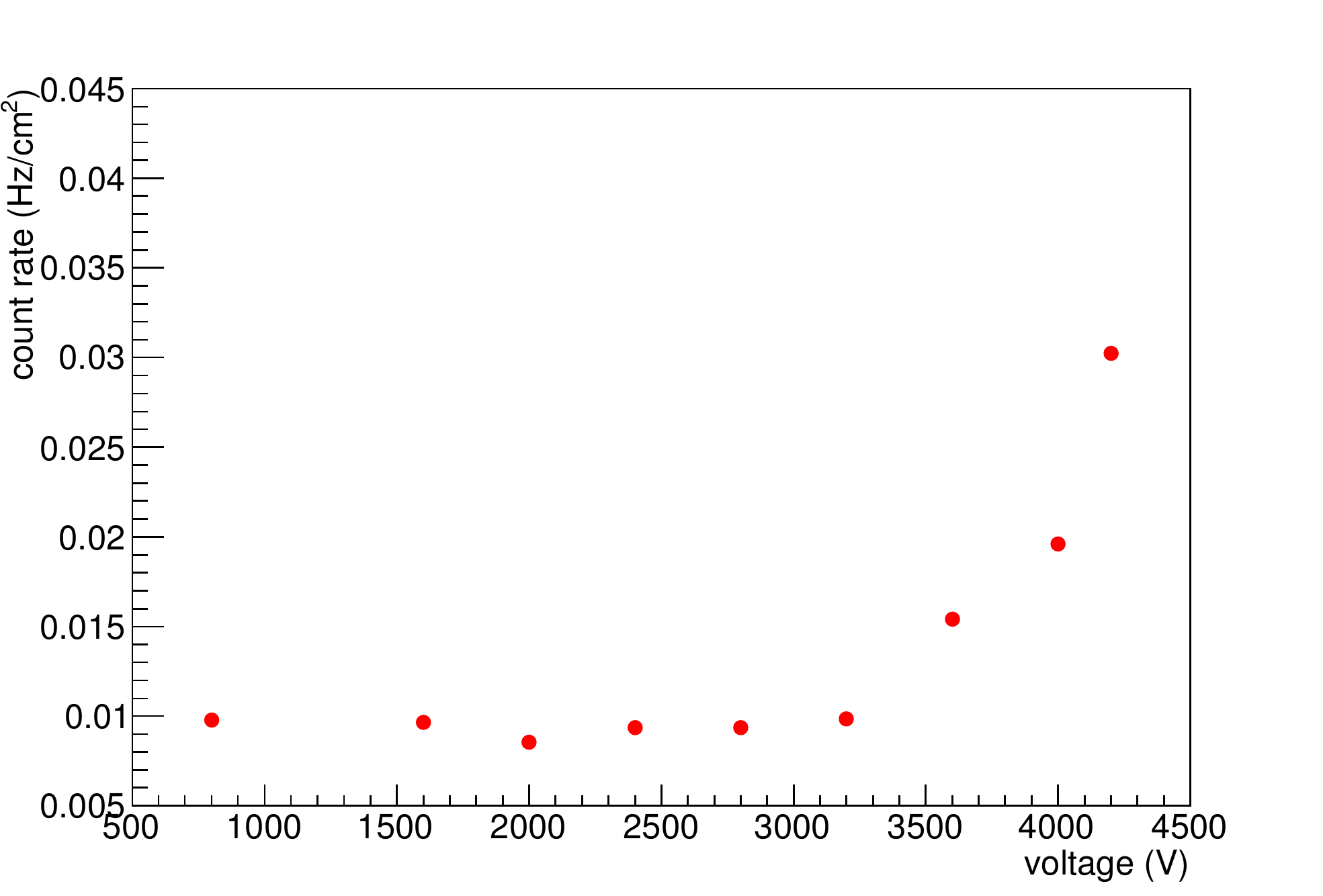}
\includegraphics[scale=0.5]{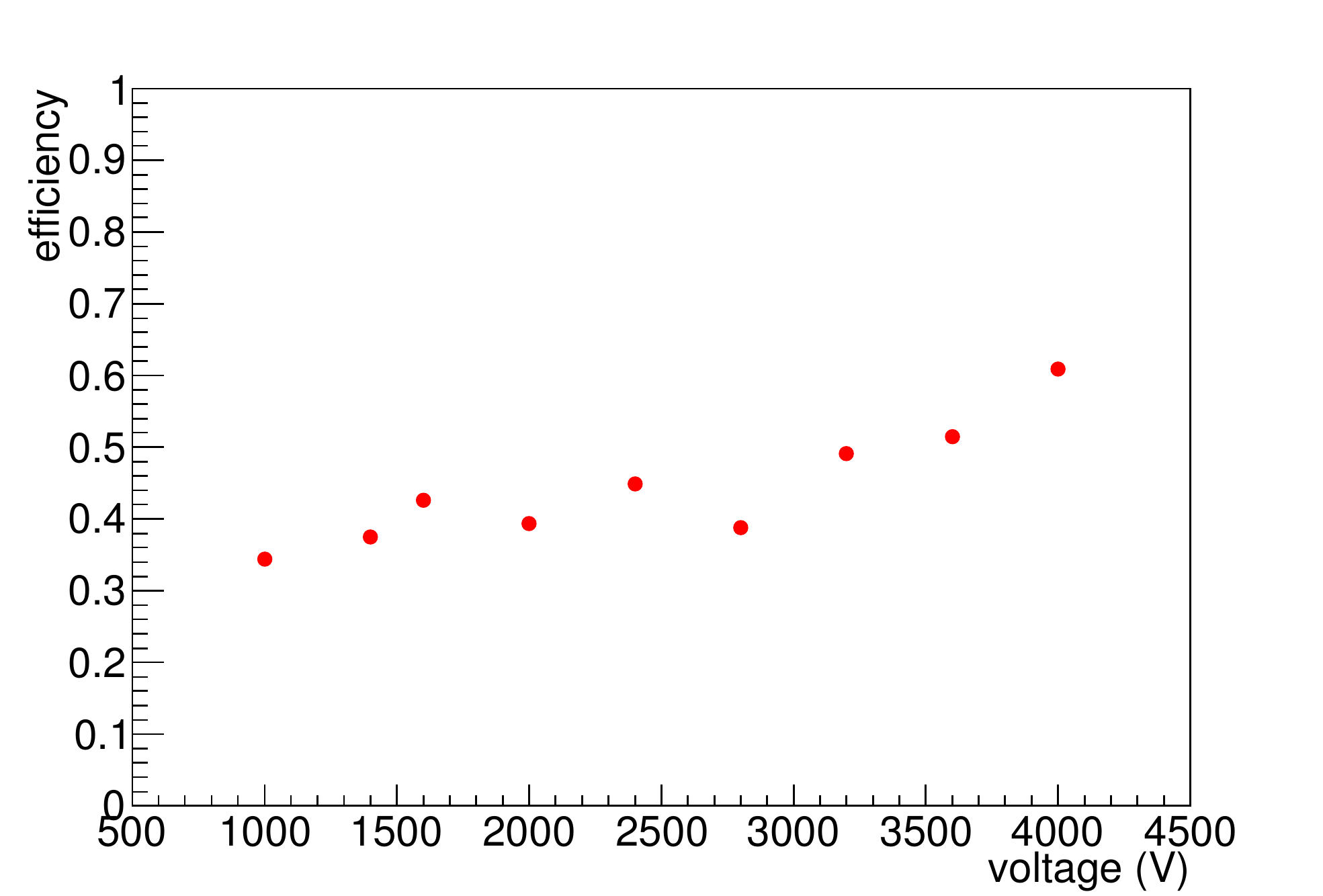}
\caption{(Left) Noise rate Vs. voltage. (Right) Efficiency Vs. voltage.}
\label{noise}
 \end{center}
\end{figure}

The three-fold signals, four-fold signals and the singles from the RPC are counted for 30~minutes for each voltage setting. The singles count rate has been divided by the area of a strip and the count rate per unit area is shown as a function of voltage in Figure~\ref{noise}. It is seen that the noise rate increases with voltage. The ratio of the four-fold count rate to the three-fold count rate is the efficiency and that as a function of voltage is shown in Figure~\ref{noise}. The efficiency increases with increasing applied voltage. At a voltage of 4~kV a typical value of efficiency and noise rate are found to be $\sim$~60\% and 0.02 Hz/cm$^2$ respectively. This value of efficiency is quite low considering a typical single gap efficiency of 90~\%. However for this detector discharges happen beyond the high voltage of 4~kV.




\section{Summary and outlook}

A single gap RPC prototype has been fabricated with very low resistive carbon-loaded PTFE plate commonly known as Teflon, to improve the rate capability. The detector has been tested in avalanche mode using 100\% R-134a as the sensitive gas. A charge sensitive preamplifier with gain 2~mV/fC and shaping time 300~ns has been used for signal collection. The V-I characteristics, variation of noise rate and efficiency as a function of voltage are studied. At a voltage of 4~kV an efficiency $\sim$~60\% is achieved. 

We have worked with a sample which has 25\% carbon-filled having a bulk resistivity of 10$^5$ $\Omega$-cm and in the second phase of R\&D a new carbon-loaded PTFE will be used with a lower carbon-filling which increases the bulk resistivity to $\sim$~10$^8$ $\Omega$-cm. This will allow us to operate the detector at relatively higher voltage. The long-term stability test of the modules are also in future plan.

\section{Acknowledgements}

The authors would like to thank Dr.~Christian~J.~Schmidt and Dr.~Walter~F.~J.~M{\"u}ller of GSI, Germany for valuable discussions and suggestions in the course of the study and providing some components. This work is partially supported by the research grant SR/MF/PS-01/2014-BI from Department of Science and Technology, Govt. of India and the research grant of CBM-MUCH project from BI-IFCC, Department of Science and Technology, Govt. of India. S. Biswas acknowledges the support of DST-SERB Ramanujan Fellowship (D.O. No. SR/S2/RJN-02/2012).


\end{document}